\PassOptionsToPackage{numbers,sort&compress}{natbib} 
\documentclass[%
preprint,
titlepage,
superscriptaddress,
nofootinbib,
 amsmath,amssymb,
 aps,
]{revtex4-2}

\usepackage{graphicx}
\usepackage{subcaption}
\usepackage{dcolumn}
\usepackage{bm}
\usepackage{threeparttable}
\usepackage{caption}
\usepackage{xcolor}
\usepackage[colorlinks=true, citecolor=blue, linkcolor=blue, urlcolor=blue]{hyperref}
\begin{document}

\preprint{}

\title{
Exponential Quintessence Model: Analytical Quantification of the Fine-Tuning Problem in Dark Energy
}

\author{Naoto Maki}
\affiliation{Department of Astronomical Science, The Graduate University for Advanced Studies (SOKENDAI), 2-21-1 Osawa, Mitaka, Tokyo 181-8588, Japan}
\affiliation{Division of Science, National Astronomical Observatory of Japan, 2-21-1 Osawa, Mitaka, Tokyo 181-8588, Japan}

\author{Kazunori Kohri}
\affiliation{Division of Science, National Astronomical Observatory of Japan, 2-21-1 Osawa, Mitaka, Tokyo 181-8588, Japan}
\affiliation{Department of Astronomy, University of Tokyo, Bunkyo-ku, Hongo, Tokyo 113-0033, Japan}
\affiliation{Department of Astronomical Science, The Graduate University for Advanced Studies (SOKENDAI), 2-21-1 Osawa, Mitaka, Tokyo 181-8588, Japan}
\affiliation{Theory Center, IPNS, KEK,
1-1 Oho, Tsukuba, Ibaraki 305-0801, Japan}
\affiliation{Kavli IPMU (WPI), UTIAS, University of Tokyo, Kashiwa, Chiba 277-8583, Japan}

\date{\today}

\begin{abstract}
  In this paper, we investigate a quintessence field with an exponential potential motivated by the suggestion of time-varying dark energy from the DESI galaxy survey. Assuming a kination epoch in the early Universe, we analytically derive constraints on initial conditions that are consistent with Big Bang Nucleosynthesis and the current dark energy density. Compared to the severe 120-digit fine-tuning
  required for dark energy to be a cosmological constant, our result
  suggests that the degree of fine-tuning is naturally relaxed by dozens
  of orders of magnitude. Furthermore, we discuss the method for
  testing this model through future observations of the gravitational
  wave background.
\end{abstract}

\maketitle

\section{Introduction}
\label{sec:Introduction}

Observations have indicated that the expansion of the Universe is
accelerating at present.  Within the framework of general relativity,
this cosmic acceleration requires dark energy, which is an unknown
energy component satisfying the condition that its equation of state
parameter $w$ is less than $-1/3$. The standard $\Lambda \mathrm{CDM}$ model achieves
acceleration through a positive cosmological constant satisfying the
equation of state parameter $w=-1$.

However, the cosmological constant requires severe fine-tuning. This
problem is known as the cosmological constant problem. For example, in
the early Universe with the Planck-scale energy, extreme 120-digit
fine-tuning is required. Furthermore, another fine-tuning problem also
exists known as the “coincidence problem”: why are the energy
densities of matter and dark energy similar in magnitude at present?
(For these fine-tuning problems, see Refs.~\cite{Sahni:2002kh,Peebles:2002gy}). 

Recently, observations of Baryon Acoustic Oscillations (BAO) by the
DESI galaxy survey have suggested a present-time variation in the
equation of state parameter of dark
energy~\cite{DESI:2024mwx,DESI:2025zgx}(see also Ref.~\cite{Tada:2024znt,Yang:2024kdo,RoyChoudhury:2025iis} for general interpretations of the DESI data). Such a present evolution of
the equation of state cannot be explained by the simple cosmological
constant. One solution to the problems associated with the
cosmological constant is to attribute acceleration to a dynamical
scalar field, called ``quintessence.'' This paper focuses on
quintessence with an exponential potential. In the early Universe, the
kinetic energy of the quintessence field dominates (called the
``kination'' epoch), while in most of the late Universe, it behaves as
a cosmological constant. Furthermore, in the present Universe, its equation of state parameter naturally fits the DESI's constraints.

For successful Big Bang Nucleosynthesis (BBN) to occur in the early
Universe, the kination epoch must end well before the temperature
decreases down to $4 \,\mathrm{MeV}$~\cite{Kawasaki:1999na,Kawasaki:2000en,Ichikawa:2005vw,deSalas:2015glj,Hasegawa:2019jsa,Barbieri:2025moq},
resulting in a radiation-dominated Universe.  Compared to the severe
120-digit fine-tuning required for dark energy as a cosmological
constant, we found that the degree of fine-tuning tends to be relaxed by dozens of orders of magnitude.  The existence of the kination epoch modifies the spectrum of the
gravitational wave background during the period.  If future
observations can detect this modification on the gravitational wave
background, it will provide evidence for this scenario.

In Sec.~\ref{sec:outline}, we outline the contents of the paper. In
Sec.~\ref{sec:Formalism}, we briefly introduce the theoretical framework for the analytical calculations, and in Sec.~\ref{sec:dynamics of scalar
  field}, we explain the dynamics of the scalar field, showing how
this model naturally fits the observations under fewer
assumptions. Sec.~\ref{sec:fine tuning problem} demonstrates the
reduced fine-tuning problem compared to the cosmological constant, and
Sec.~\ref{sec:primordial gravitational wave} describes the method for
testing this model through the future observations of the
gravitational wave background. Sec.~\ref{sec:Conclusion} is devoted to
the conclusion.

Throughout this paper, we adopt units where $c=\hbar=k_{\rm B}=1$.

\section{Outline}
\label{sec:outline}

In this paper, we focus on a quintessence model with the exponential
potential of a scalar field $\phi$, which gives
$V(\phi)=V_0e^{-\lambda\phi/m_{\mathrm{pl}}}$~\cite{Ferreira:1997hj,Copeland:1997et},
where $V_0$ is positive constant with $m_{\mathrm{pl}}$ being the reduced
Planck mass ($=2.4 \times 10^{18}$~GeV), and $\lambda$ is a positive
dimensionless parameter. So far, such quintessence with the exponential potential has
been studied in detail by analyzing the dynamical
system~\cite{Halliwell:1986ja,Burd:1988ss,Copeland:1997et,Urena-Lopez:2011gxx,Tamanini:2014mpa,vandenHoogen:1999qq,Gosenca:2015qha,Boehmer:2010jqg,SavasArapoglu:2017pyh}
(see also~\cite{Bahamonde:2017ize,Copeland:2006wr} for
reviews). Regarding the origin and the properties of $\lambda$,
numerous arguments have been done, e.g., in terms of predictions from
string theory~\cite{Obied:2018sgi}, scale-invariant gravity~\cite{Hong:2025tyi}, multi-field
extensions~\cite{Liddle:1998jc,Malik:1998gy,Chiba:2014sda,Alestas:2025syk}, 
theoretical upper bounds~\cite{Andriot:2024jsh}, and observational
constraints~\cite{Bhattacharya:2024hep,Ramadan:2024kmn,Akrami:2025zlb,Bayat:2025xfr,Pourtsidou:2025sdd}.

In the exponential quintessence models, we can find several attractor
solutions where the scalar field evolves while maintaining a constant
ratio of energy density to that of matter or radiation (see
TABLE~\ref{tab:fixedpoint} in the next section). Such a scalar field
solution can partly resolve both the fine-tuning problems (i.e., the
cosmological constant problem and the coincidence problem). However,
the problem is that the scalar field tends to be still subdominant and
can not produce the acceleration of the cosmic expansion. This is
because the energy density of the scalar field is just following to be
proportional to the existing background fluid density of matter or
radiation in the attractor solutions.

Even among the exponential potential models, however, we can achieve the current cosmic acceleration, by constructing solutions that do not follow the attractor from the early Universe and only approach it intermediately near the present epoch~\cite{Kolda:2001ex}. In this paper, we focus on such scenarios. While such scenarios can account for the cosmic acceleration, the absence of attractor behavior in the past implies that the current state once again depends on initial conditions, causing the fine-tuning problem to re-emerge.

Regarding the fine-tuning problem in the exponential quintessence
model, the parameter space that fit both the BBN and the current cosmic acceleration were explored in ~\cite{Franca:2002iju}.
By fixing the initial value of the kinetic
energy-density parameter to be $\Omega_{\mathrm{kin,i}}=0.00125$ at
$z=10^{13}$, they showed that realistic cosmological solutions are achievable within the range $0 \leq \lambda \lesssim 1.7$.

Despite those past works in the exponential quintessence models, the degree of fine-tuning required for the initial condition of the
density parameter of the potential energy $\Omega_{\mathrm{pot,i}}$
has not been determined. Unlike the case for the cosmological
constant, which requires fine-tuning of the order of
$\Omega_{\mathrm{pot,pl}}\sim 10^{-120}$ at the Planck time, the
degree of fine-tuning required for quintessence is non-trivial due to
the time evolution of its energy density. However, our purpose is to
analytically quantify the degree of fine-tuning required for the
exponential quintessence. Specifically, we consider a scenario
inspired by string theory~\cite{Cicoli:2023opf}. In this model, the quintessence field $\phi$ couples with
the inflaton field, and it undergoes kination just after the end of
primordial inflation by rolling down the exponential potential in the
early Universe and behaves as dark energy in the late Universe. 

In Sec.~\ref{sec:dynamics of scalar field} we show that the scalar field remains effectively frozen after the transition from the kination epoch to the radiation-dominated epoch, and that the current dark energy density is characterized by the potential energy at this
transition. In addition we demonstrate that this model can explain time-varying equation of state suggested by DESI's second data release without additional fine-tuning.

In Sec.~\ref{sec:fine tuning problem}, we derive the initial condition of
$\Omega_{\mathrm{pot},\mathrm{pl}}$ to satisfy the constraints from both
the BBN and the observed energy-scale of dark energy to be
\begin{align}
  \Omega_{\mathrm{pot},\mathrm{pl}} \lesssim 10^{-120} 10^{21\sqrt{6}\lambda} \notag.
\end{align}
This inequality quantifies fine-tuning required for exponential
quintessence. It indicates that the degree of fine-tuning can be
naturally relaxed relative to the cosmological constant for a non-zero
value of $\lambda$.

In Sec.~\ref{sec:primordial gravitational wave}, we further calculate
the spectrum of the primordial gravitational wave background amplified
by entering the horizon during the kinetic epoch under this setup.  We
show that future observations of the gravitational waves will test
this scenario.

\section{Formalism}
\label{sec:Formalism}

We assume a four-dimensional spatially flat homogeneous and isotropic Universe. The Friedmann-Lema\^itre-Robertson-Walker (FLRW) metric for the flat spacetime is given by
\begin{align}
  ds^2=-dt^2 +a(t)^2 \left[ dr^2+r^2(d\theta^2+\sin^2 \theta d\phi^2) \right],
\end{align}
where we adopt the normalization of the scale factor $a$ to be unity
at the present time. In addition to radiation and matter as the
standard energy components, we consider a real scalar field
$\phi$. Then, the energy density and pressure of $\phi$ are given by
\begin{align}
  &\rho_\phi=\frac{1}{2}\dot{\phi}^2+V(\phi),\\
  &p_\phi=\frac{1}{2}\dot{\phi}^2-V(\phi),
\end{align}
respectively. Here $\dot{\phi} = \frac{d \phi}{dt}$ means the
derivative of $\phi$ with respect to cosmic time $t$. Equation of state
parameter $w_\phi\equiv p_\phi/\rho_\phi$ satisfies
$-1\leq w_\phi \leq1 $. Here, we consider exponential potential
\begin{align}
  V(\phi)=V_0 e^{-\lambda \phi/m_{\mathrm{pl}}},
\end{align}
where $\lambda$ and $V_0$ are positive constants and $m_{\mathrm{pl}}$ is the reduced Planck mass.  The Einstein equation and the equation of motion for the scalar field are given by 
\begin{align}
  &3m_{\mathrm{pl}}^2 H^2 = \frac{1}{2}\dot{\phi}^2+V(\phi)+\rho_m+\rho_r\label{Friedmann equation},\\
  &2m_{\mathrm{pl}}^2 \dot{H}=-\dot{\phi}^2-\rho_m-\frac{4}{3}\rho_r\label{acceleration equation},\\
  &\ddot{\phi}+3H\dot{\phi}+V_{,\phi}=0,\label{scalar field EOM}
\end{align}
respectively, where $V_{,\phi}=\frac{dV}{d\phi}$. Furthermore,
following the notation of \cite{Bhattacharya:2024hep} we define the
dimensionless variables as follows:
\begin{align}
  \Omega_m\equiv\frac{\rho_m}{3m_{\mathrm{pl}}^2H^2},\qquad x\equiv\frac{\dot{\phi}}{\sqrt{6}m_{\mathrm{pl}} H},\qquad y\equiv\frac{\sqrt{V}}{\sqrt{3}m_{\mathrm{pl}} H},\qquad u\equiv\frac{\sqrt{\rho_r}}{\sqrt{3}m_{\mathrm{pl}}H}.
  \label{dimensionless parameter}
\end{align}
The squares of $x,y,u$ correspond to the density parameter of the
scalar field kinetic energy $\Omega_{\mathrm{kin}}\equiv x^2$, the
potential energy density parameter $\Omega_{\mathrm{pot}}\equiv y^2$,
and the radiation density parameter $\Omega_r\equiv u^2$,
respectively. Here, denoting the derivative with respect to the
e-folding number $N=\ln a$ by a prime, $'=\frac{d}{dN}$, we obtain the following
relations for $x,y$ and $u$:

\begin{align}
  &x'=\frac{1}{2}x(-3+u^2+3x^2-3y^2)+\sqrt{\frac{3}{2}}\lambda y^2\label{x_dif},\\
  &y'=\frac{1}{2}y(3+u^2+3x^2-3y^2)-\sqrt{\frac{3}{2}}\lambda xy\label{y_dif},\\
  &u'=\frac{1}{2}u(-1+u^2+3x^2-3y^2).
\end{align}
The fixed points of this system of equations are given in Table~\ref{tab:fixedpoint}.
\begin{table}[htbp]
\centering
\renewcommand{\arraystretch}{1.5} 
\resizebox{\textwidth}{!}{
\begin{tabular}{|c|c|c|c|c|} 
\hline

&Fixed points $(x,y,u)$ &Eigenvalues & Existence & Stability \\
\hline \hline

$A_1$ &$(+1,0,0)$ &  $\left( 3,3- \lambda\sqrt{\frac{3}{2}},1 \right)$ &$\forall\lambda$ & \begin{tabular}[c]{@{}c@{}}  Unstable for  $\lambda\le\sqrt{6}$ \\ \small Saddle for $\lambda>\sqrt{6}$\end{tabular} \\
\hline

$A_2$ &$(-1,0,0)$ &  $\left( 3,3+ \lambda\sqrt{\frac{3}{2}},1 \right)$ &$\forall\lambda$ & \begin{tabular}[c]{@{}c@{}}  Unstable 
\end{tabular} \\
\hline

$B$ &$(0,0,0)$ & $\left( -\frac{3}{2},\frac{3}{2},-\frac{1}{2} \right)$ &  $\forall\lambda$ & Saddle \\
\hline

$C$ &$\left( \frac{\lambda}{\sqrt{6}},\pm\frac{\sqrt{6-\lambda^{2}}}{\sqrt{6}},0 \right)$ & $\left( \frac{\lambda^2}{2}-3,\lambda^2-3,\frac{\lambda^2}{2}-2 \right)$ & $\lambda<\sqrt{6}$  & \begin{tabular}[c]{@{}c@{}}Stable for $\lambda<\sqrt{3}$\\ Saddle for $\lambda>\sqrt{3}$\end{tabular} \\
\hline

$D$ &$\left( \frac{1}{\lambda}\sqrt{\frac{3}{2}},\pm\frac{1}{\lambda}\sqrt{\frac{3}{2}},0 \right)$ &  $\left( -\frac{3(\lambda+\sqrt{24-7\lambda^2})}{4\lambda}, -\frac{3(\lambda-\sqrt{24-7\lambda^2})}{4\lambda} ,-\frac{1}{2}\right)$& $\lambda>\sqrt{3}$ & Stable \\
\hline

$E$ &$(0,0,\pm1)$ & $(-1,2,1)$ &$\forall\lambda$ & Saddle \\
\hline

$F$ &$\left( \frac{1}{\lambda}\sqrt{\frac{8}{3}},\pm\frac{2}{\lambda\sqrt{3}},\pm\sqrt{1-\frac{4}{\lambda^{2}}} \right)$  &$\left( -\frac{\lambda+\sqrt{64-15\lambda^2}}{2\lambda},-\frac{\lambda-\sqrt{64-15\lambda^2}}{2\lambda},1\right)$& $\lambda>2$ & Saddle \\
\hline
\end{tabular}
}
\caption{Each row displays the fixed point, the eigenvalues of Jacobian, the condition of $\lambda$ for the existence, and the stability. Note that we consider the cases only for $\Omega_k=0$, which were studied in~\cite{Andriot:2024jsh,Bhattacharya:2024hep}}
\label{tab:fixedpoint}
\end{table}
The points $A$, $B$, and $E$ correspond
to epochs dominated by the kinetic energy of the scalar field, matter,
and radiation, respectively. $D$ and $F$ correspond to
the scaling solutions, in which the scalar field mimics the evolution
of other energy components (i.e., matter and radiation). $C$
corresponds to an epoch dominated by the scalar field where its
kinetic energy and potential energy evolve similarly.  The asymptotic
state of the Universe depends on $\lambda$. When $\lambda < \sqrt{3}$,
the Universe approaches the stable point $C$ where
$w_\phi=w_{\mathrm{eff}}=-1+\frac{\lambda^2}{3}$.  Conversely, when
$\lambda \ge \sqrt{3}$, the point $D$ is stable where
$w_\phi=w_{\mathrm{eff}}=0$.

\section{Dynamics of Scalar Field}
\label{sec:dynamics of scalar field}

In this section, we outline the cosmological evolution of the scalar
field. The dynamics can be broadly categorized into three phases: (1)
a kination phase immediately after the end of inflation, (2) a freezing phase
where the field remains nearly constant during radiation domination
and/or matter domination, and (3) an attractor phase where the field
resumes rolling down the potential. The present Universe corresponds
to the transition from the end of phase (2) to the beginning of phase
(3).

In this study, we consider a scenario where the field value of $\phi$
enters a region described by an exponential potential after the end of
inflation. By rolling down this potential, inflation ends, and the field enters a
kination phase of $\phi$. Finally, it behaves as a dark energy. This concept may include models of the quintessential inflation~\cite{Agarwal:2018obp,deHaro:2021swo,Feng:2002nb}. However, the quintessential field can be different from the inflaton field.
This setup is
motivated by superstring theory where moduli fields typically have
exponential potentials. The presence of a kination epoch provides a
solution to the overshoot problem of moduli \cite{Cicoli:2023opf}. We
assume that radiation is generated by the decay of another spectator
field, such as a curvaton (various other mechanisms for radiation
production are possible; see appendix A of
\cite{Gouttenoire:2021jhk}).

Fig.~\ref{fig:rho_evolution} shows the evolution of energy density of
realistic solutions for $\lambda=1$ or $\sqrt{3}$.
\begin{figure}[htbp]
  \centering
  \begin{minipage}{.45\linewidth}
    \centering
    \includegraphics[width=\linewidth]{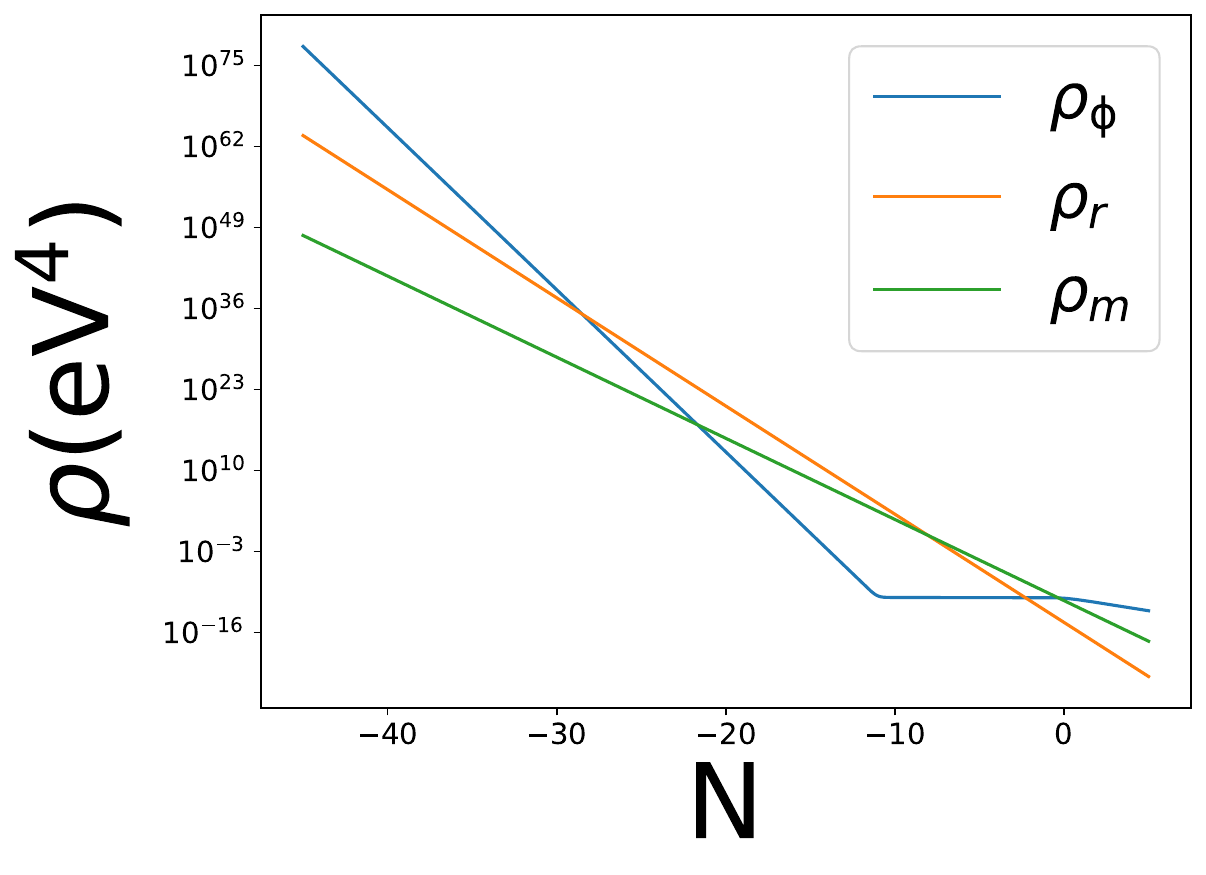}
    \subcaption{}
  \end{minipage}
  \hspace{1mm}
  \begin{minipage}{.45\linewidth}
    \centering
    \includegraphics[width=\linewidth]{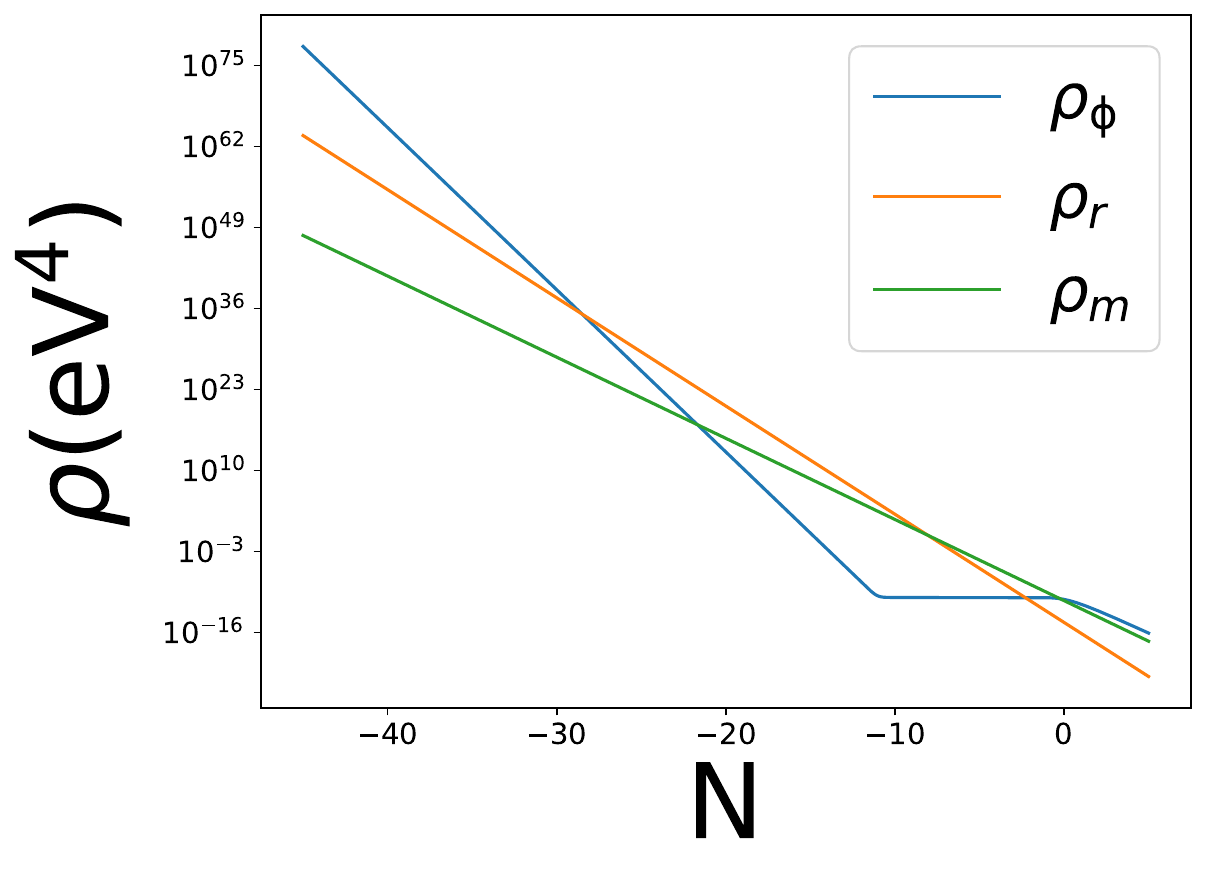}
    \subcaption{}
  \end{minipage}
  \vspace{3mm}
  \begin{minipage}{.45\linewidth}
    \centering
    \includegraphics[width=\linewidth]{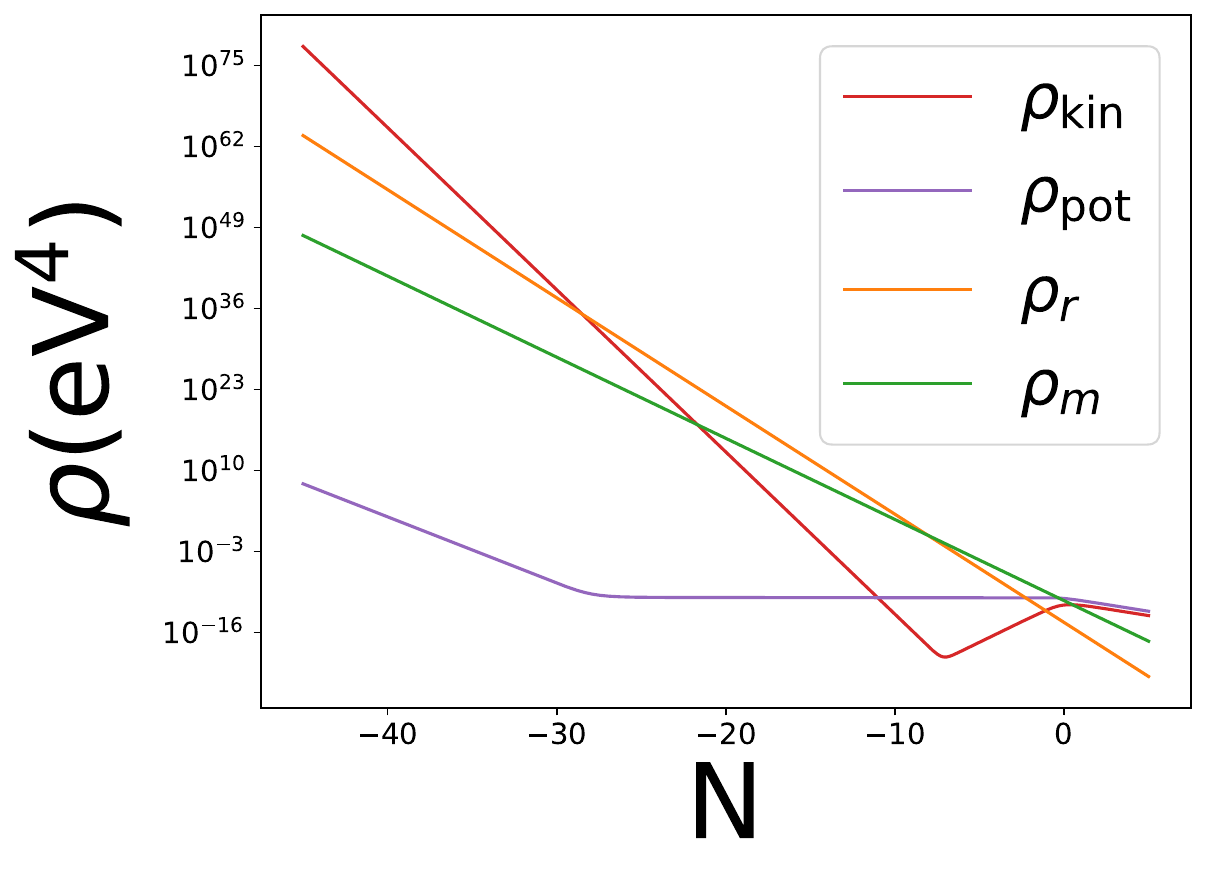}
    \subcaption{}
  \end{minipage}
  \hspace{1mm}
  \begin{minipage}{.45\linewidth}
    \centering
    \includegraphics[width=\linewidth]{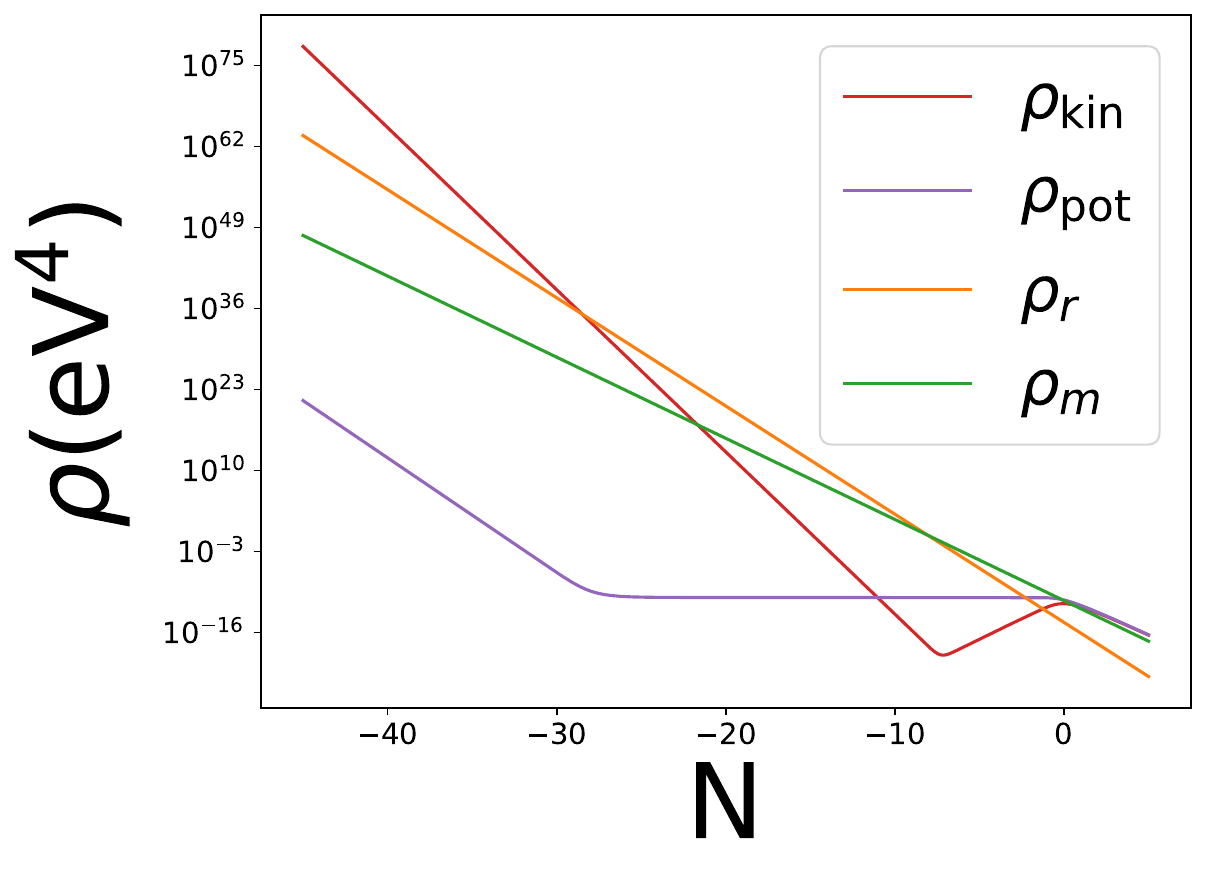}
    \subcaption{}
  \end{minipage}
  \caption{The curves represent the energy densities of total scalar field (blue), radiation (orange),  matter (green), kinetic term
    (red) and potential term (purple), respectively. The left
    column, i.e., (a), and (c), corresponds to the cases for
    $\lambda=1.0$, while the right column, i.e, (b) and (d),
    corresponds to the cases for $\lambda=\sqrt{3}$. In the top
    panels (a) and (b), we show the total (kinetic + potential) energy
     density of $\phi$. On the other hand, in the bottom panels (c) and (d), we show the kinetic and potential energies of $\phi$ separately.}
  \label{fig:rho_evolution}
\end{figure}
The upper panels show the kinetic and potential energies of $\phi$
separately, while the lower panels present the total (kinetic +
potential) energy density of $\phi$. In the following subsection, we
analytically analyze the behavior of the scalar field in each epoch.
\subsection{Kination Phase}
First, we consider the kination phase. In this phase, the potential term
in the equation of motion is negligible. Since the scale factor
evolves as $a \propto t^{1/3}$, the Hubble parameter is given by
$H=1/3t$, and the equation of motion becomes,
\begin{align}
  \ddot{\phi}+\frac{1}{t}\dot{\phi}=0.
\end{align}
This leads to 
\begin{align}
    \dot{\phi}=\frac{C}{t}
\end{align}
where $C$ is a constant and solved to be $C=\pm\sqrt{\frac{2}{3}}m_{\mathrm{pl}}$. Therefore, $\phi$ is given by
\begin{align}
    \phi=\phi_i\pm\sqrt{6}m_{\mathrm{pl}}(N-N_i), \label{phi_kination}
\end{align}
where $N_i$ and $\phi_i$ are initial e-folding number and initial
value of $\phi$ at initial time, respectively. For the case
$\dot{\phi}>0$, the potential is represented by
\begin{align}
  V=V_0 e^{\lambda \phi_i/m_\mathrm{pl}}e^{-\sqrt{6}\lambda(N-N_i)}.
\end{align} 
Thus, the potential energy scales as V $\propto
a^{-\lambda\sqrt{6}}$.
This indicates that for $\lambda> \sqrt{6}$, the potential energy
decreases more rapidly than kinetic energy, whereas for
$\lambda<\sqrt{6}$, it decreases more slowly than the kinetic energy.
\subsection{Freezing Phase}
\begin{figure}[htbp]
  \centering
  \begin{minipage}{.45\linewidth}
    \centering
    \includegraphics[width=\linewidth]{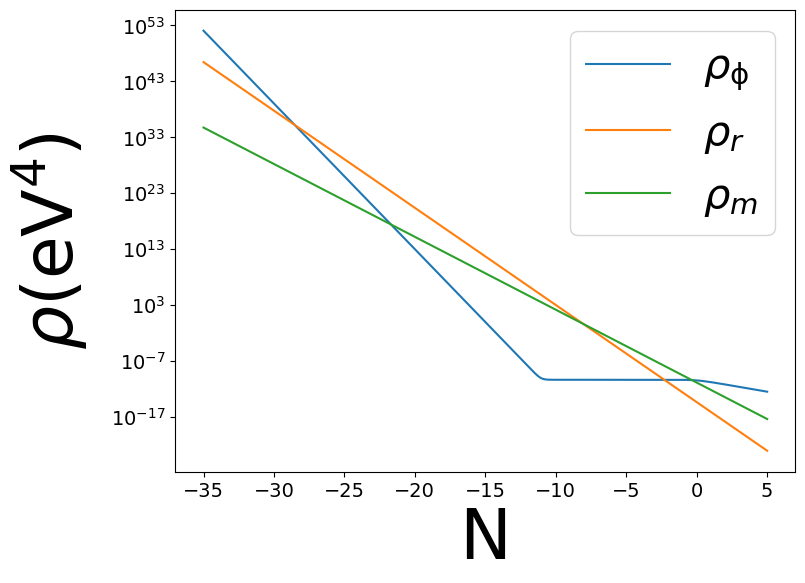}
    \subcaption{}
  \end{minipage}
  \hspace{1mm}
  \begin{minipage}{.45\linewidth}
    \centering
    \includegraphics[width=\linewidth]{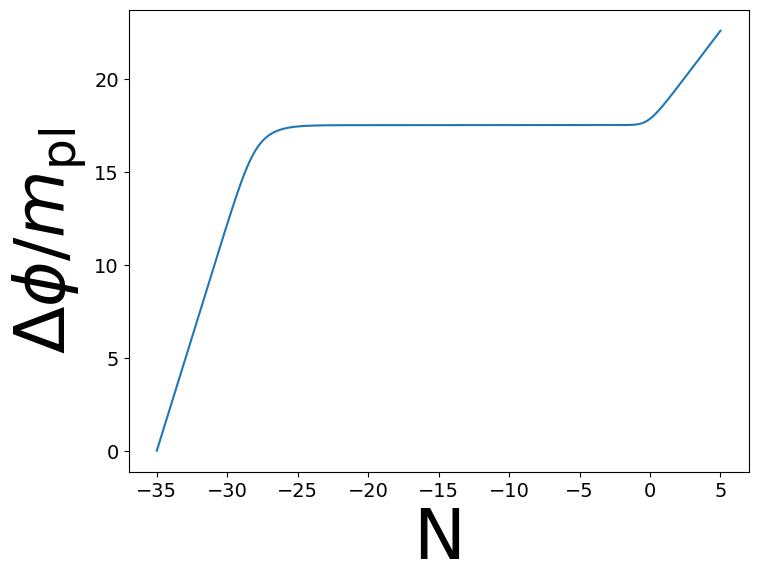}
    \subcaption{}
  \end{minipage}
  \caption{(a) Time evolution of the energy density for $\lambda=1$, using the same initial conditions as the left column of Fig. \ref{fig:rho_evolution}. (b) The excursion of the scalar field value for $\lambda=1$. The plots indicate that the field becomes frozen once radiation domination begins and remains fixed until around the present epoch.}
  \label{fig:field excursion}
\end{figure}

Since the kinetic energy decreases more rapidly ($\propto a^{-6}$)
than radiation ($\propto a^{-4}$), the scalar field eventually becomes
subdominant. Once the scalar field becomes subdominant
($\Omega_\phi\ll 1$), the value of the scalar field becomes
effectively constant. This is called the freezing phase. This freezing
behavior can be roughly estimated by evaluating the excursion per
Hubble time:
\begin{align}
  \frac{\dot{\phi}}{H}=\sqrt{3\Omega_\phi(1+w_\phi)}m_{\mathrm{pl}},
  \label{excursion_per_Hubble_time}
\end{align}
where we used $\dot{\phi}=\sqrt{\rho_\phi(1+w_\phi)}$. Therefore, the
field excursion is negligible as long as $\Omega_\phi \ll 1$.
Consequently, the field effectively freezes after radiation domination
begins until around the present epoch where $\Omega_\phi\sim
1$.
Fig.~\ref{fig:field excursion} illustrates that the scalar field
becomes effectively frozen from radiation domination to around the
present epoch.
It is important to note that although the freezing of the scalar field
implies a constant potential energy, the kinetic energy remains highly
dynamic during this freezing epoch as shown in
Fig.~\ref{fig:rho_evolution}. To see this quantitatively, it is
advantageous to focus on the evolution of the dimensionless
variables. Assuming radiation domination, the conditions
$V\simeq\mathrm{const}$ and $H=1/2t$ imply $y=y_i e^{2(N-N_i)}$.
Furthermore, since $x,y\ll 1$ and $u\simeq 1$, Eq.~\eqref{x_dif} can be
approximated to be
\begin{align}
    x'\simeq -x+\sqrt{\frac{3}{2}}\lambda y_i^2 e^{4(N-N_i)}.
\end{align}
This equation can be solved analytically:
\begin{align}
  x=A e^{-(N-N_i)}+B e^{4(N-N_i)},
  \label{x_sol}
\end{align} 
where A and B are constants determined by the initial conditions. The
first term of Eq.\eqref{x_sol} represents the decay mode, and on the
other hand, the second term represents the growth mode. First of all, the
first term is much larger than the second term, and the kinetic energy drops as
$\dot{\phi}^2/2\propto x^2H^2\propto e^{-2N}e^{-4N}\propto a^{-6}$.
Eventually, the growing term surpasses the decaying term, and the
kinetic energy begins to increase.

This transition corresponds to a shift in the dominant term of the
equation of motion. In the early freezing stage where the decaying
term dominates, the potential gradient $V_{,\phi}$ is
negligible. Assuming radiation domination ($H=1/2t$), the equation of
motion is approximated as
\begin{align}
    \ddot{\phi}+\frac{3}{2t}\dot{\phi}=0.
\end{align}
Then, the solution is given by 
\begin{align}
    \phi=\phi_i +\frac{\dot{\phi}_i}{H_i}\left( 1-e^{-(N-N_i)} \right),
\end{align}
where $\phi_i$, $\dot{\phi_i}$ and $H_i$ are the field value, the
field velocity, and the Hubble parameter at the time $N_i$,
respectively.  This solution explicitly shows that the field
effectively freezes due to the Hubble friction after radiation
domination begins. Eventually, the potential term becomes no longer
negligible in the equation of motion, which corresponds to the epoch
when the growth mode begins to dominate.\footnote{In this growing
  regime, all terms in the equation of motion become comparable to
  each other, and their relative fractions become
  constant~\cite{Linder:2006sv,Urena-Lopez:2011gxx}. Specifically, if
  the Universe is in radiation domination during this growth phase,
  the ratio is given by $\frac{\ddot{\phi}}{3H\dot{\phi}}= \frac23$
  and $\frac{V_{,\phi}}{3H\dot{\phi}}=-\frac53$. Similarly in the
  matter domination, the terms keep the ratio:
  $\frac{\ddot{\phi}}{3H\dot{\phi}}= \frac12$ and $\frac{V_{,\phi}}{3H\dot{\phi}}=-\frac32$.} 

\begin{figure}[htbp]
  \centering
  \includegraphics[width=0.8\columnwidth]{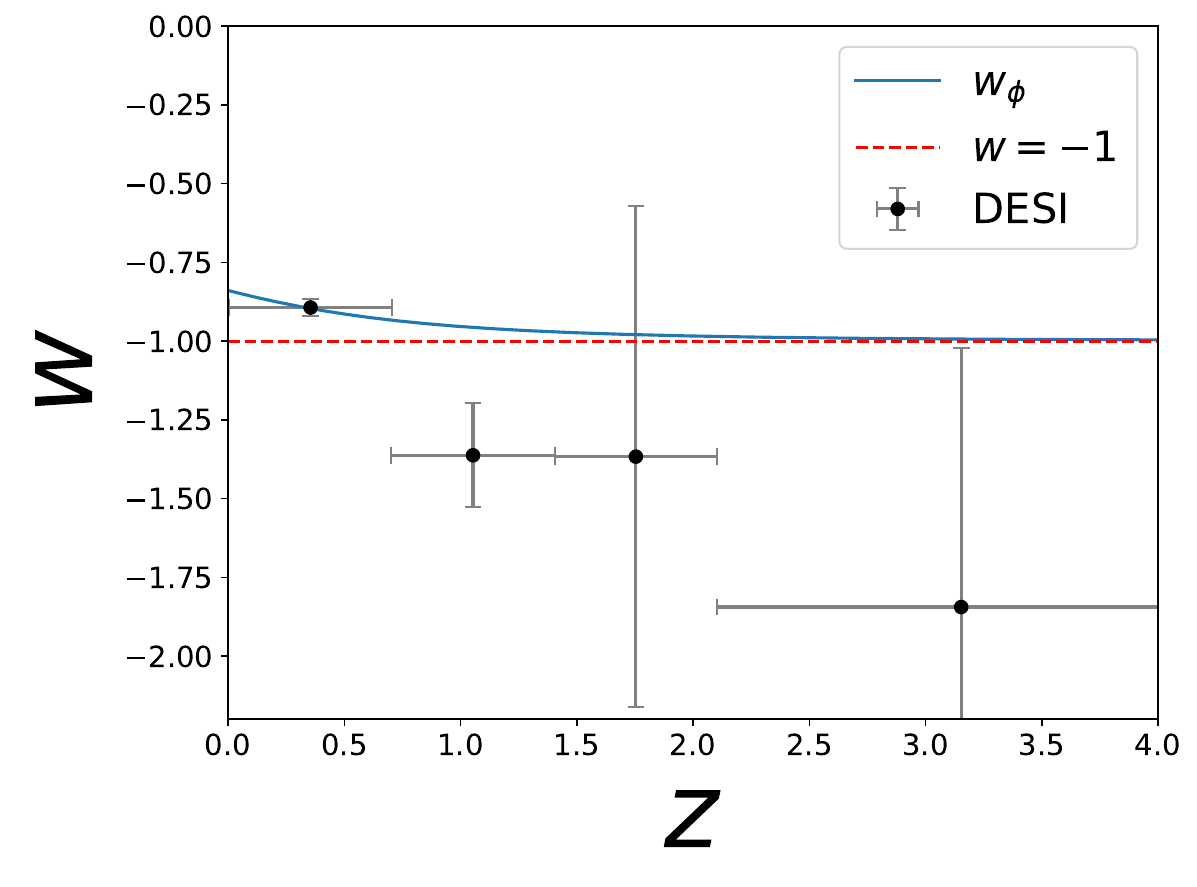}
  \caption{Comparison of evolution of $w_\phi$ for $\lambda=1$ with the data points from DESI's second data release \cite{DESI:2025zgx}. Blue solid line represents the evolution of $w_{\phi}$ using the same initial condition of $\lambda=1$ case in Fig.~\ref{fig:rho_evolution}. The red dashed line shows the cosmological constant ($w=-1$). The black points, $1\sigma$ error bar (vertical bars) and bin width (horizontal bar) are taken from Fig.12 of \cite{DESI:2025zgx}.  }
  \label{fig:w}
\end{figure}
As the energy density of the scalar field becomes non-negligible relative to the total energy of the Universe, the field begins to roll down the potential as can be seen from Eq.~\eqref{excursion_per_Hubble_time}. Fig.~\ref{fig:w} shows the evolution of the equation of state parameter $w_{\phi}$ for $\lambda=1$, compared with observational data points from DESI's second data release \cite{DESI:2025zgx}. The dynamics of the scalar field causes the equation of state parameter to depart from its behavior as an effective cosmological constant ($w_{\phi}\simeq-1)$. Consequently, in this model, the time-varying equation of state in dark energy suggested by DESI arises naturally from the field dynamics without any fine-tuning for the timing. This is because the transition to the dark energy dominated epoch inevitably coincides with the beginning of the time variation of $w_{\phi}$.

Finally, the field
settles into its asymptotic state, which is either the scaling
solution with matter or the domination of scalar field. The detail of
this final stage does not affect the degree of fine-tuning for the
initial conditions, and thus we can omit it. The most important
consequence of the analysis in this section is that, due to the field
freezing, the potential energy of the scalar field remains effectively
constant from the end of the kination phase to the vicinity of the
present epoch as shown in Fig.~\ref{fig:rho_evolution} and
Fig.~\ref{fig:field excursion}. This means that the current dark energy density is characterized by the potential energy at the transition from
the kination epoch to the radiation-dominated epoch. This feature plays a crucial role in
the analytical argument of fine-tuning discussed in the next section.

\section{Fine-tuning problem}
\label{sec:fine tuning problem} 
In this section, we first discuss the meaning of the condition
$T^{\mathrm{eq}}_{\mathrm{kin-r}}> 4\,\mathrm{MeV}$ derived
in Sec.~\ref{sec:BBN constraint}, where
$T^{\mathrm{eq}}_{\mathrm{kin-r}}$ is the transition temperature from
the kination phase to radiation domination. Next, in
Sec.~\ref{sec:constraint on initial condition} we derive
constraints on the initial conditions from the BBN, which is the main
subject of this paper.

\subsection{Constraints from Big Bang Nucleosynthesis}
\label{sec:BBN constraint}
In this subsection, from the requirement for the successful BBN, we
derive the condition for the end of the kinetic epoch where the
kinetic energy falls below the radiation energy.  As an initial
condition for the success of the BBN, the Universe should be in the
radiation dominated epoch to thermalize the background particles
through both weak and electromagnetic interactions~\footnote{
The
contents of the background particles are (1) photon ($\gamma$), (2)
electron and positron ($e^-$ and $e^+$), (3) three neutrinos
($\nu_e$,$\nu_{\mu}$,$\nu_{\tau}$ and their antiparticles), and (4)
nuclei (proton $p$ and neutron $n$). 
} 
In the standard BBN, the
decoupling time of the following weak reactions,
\begin{align}
  &n + \nu_e \leftrightarrow p + e^-, \notag\\
  &n + e^+ \leftrightarrow  p + \bar{\nu}_e, \notag\\
  &n  \leftrightarrow p+ e^- + \bar\nu_e,
  \label{pn}
\end{align}
can be estimated using the Gamow criterion $\Gamma=H$ with the reaction
rate $\Gamma$. During the radiation-dominated epoch, the Hubble
parameter is estimated to be
\begin{align}
  H\sim \frac{T^2}{m_{\mathrm{pl}}}
  \label{H-T relation}
\end{align}
On the other hand, according to the theory of weak interaction, the rate of the reactions
in Eq.~\eqref{pn} is simply given by
\begin{align}
  \Gamma\sim G_F^2 T^5 \sim \frac{T^5}{m_W^4},
\end{align}
where $G_F$ is the Fermi coupling constant and $m_W$ is the W boson
mass ($m_W\sim 80.4\mathrm{GeV}$). Therefore, the decoupling
temperature $T_{\mathrm{SD}}$ in the standard BBN is roughly estimated to be
\begin{align}
  T_{\mathrm{SD}}\sim \left( \frac{m_W^4}{m_{\mathrm{pl}}} \right)^{1/3}\sim {\cal O}(1)~\mathrm{MeV}.
\end{align}
When the reactions are in thermal equilibrium for
$T>T_{\mathrm{SD}}$, the neutron-to-proton ratio follows the Boltzmann
distribution. Thus, the ratio at decoupling is expressed by
\begin{align}
  \frac{n_{n}}{n_{p}}=e^{-\frac{Q_{np}}{T_{\mathrm{SD}}}}\sim 0.2,
  \label{boltzmann}
\end{align}
where the $Q$-value of the reaction is
$Q_{np}=m_n-m_p \simeq 1.29$~MeV with $m_n$ and $m_p$ being the
neutron and proton masses, respectively. After the decoupling of weak
interaction, the ratio decreases only due to free decay of neutrons
until the cosmic time $t\sim 300$~sec which gives
$n_{n}/n_{p}\sim 1/7$. This ratio partly determines the main outcome
of the BBN, especially the primordial value of the mass fraction of
$^4$He, which is estimated to be
$Y_p = \frac{2 n_{n}/n_{p}}{1 + {n_{n}}/{n_{p}}} \sim 1/4$.

In terms of the initial condition for the successful BBN, we
conservatively take a more concrete value than it, e.g., for the
minimum requirement to thermalize the background particles including
the neutrinos, we need $T > T_{\mathrm{SD}}\sim 4~\mathrm{MeV}$ in
the standard
BBN~\cite{Kawasaki:1999na,Kawasaki:2000en,Ichikawa:2005vw,deSalas:2015glj,Hasegawa:2019jsa,Barbieri:2025moq}.

However, the existence of a kination phase modifies the temperature
dependence of the Hubble parameter. Since $\rho_\phi \propto a^{-6} \propto T^{6} $,
we can express the energy density in terms of the transition
temperature $T^{\mathrm{eq}}_{\mathrm{kin-r}}$ by
\begin{align}
  \rho_\phi\sim \left( T^{\mathrm{eq}}_{\mathrm{kin-r}} \right)^4\left( \frac{T}{T^{\mathrm{eq}}_{\mathrm{kin-r}}} \right)^6=\frac{T^6}{\left( T^{\mathrm{eq}}_{\mathrm{kin-r}} \right)^2}.
\end{align}
Then, the Hubble parameter is expressed by
\begin{align}
  H\sim \frac{\sqrt{\rho_\phi}}{m_{\mathrm{pl}}}\sim \frac{T^3}{T^{\mathrm{eq}}_{\mathrm{kin-r}}m_{\mathrm{pl}}}.
\end{align}
By solving $\Gamma=H$ for the modified decoupling temperature $T'$, we obtain
\begin{align}
  T'\sim \sqrt{\frac{m_W^4}{T^{\mathrm{eq}}_{\mathrm{kin-r}}m_{\mathrm{pl}}}}\sim T_{\mathrm{SD}} \sqrt{\frac{T_{\mathrm{SD}}}{T^{\mathrm{eq}}_{\mathrm{kin-r}}}}.
\end{align}
This implies that if
$T^{\mathrm{eq}}_{\mathrm{kin-r}}<T_{\mathrm{SD}}$, the decoupling
temperature $T'$ exceeds $T_{\mathrm{SD}}$. Such a higher decoupling
temperature results in an increased neutron fraction via
Eq.~\eqref{boltzmann}, which gives a larger $^4$He abundance very
sensitively ($Y_p \gg 1/4$) and is inconsistent with observations
($Y_p =0.2458 \pm 0.0026~(95\%)$~C.L.~\cite{Aver:2026dxv}, see
also~\cite{Yanagisawa:2025mgx}).  From a conservative standpoint, if
the transition occurs earlier than the decoupling time of the standard
BBN (i.e. $T^{\mathrm{eq}}_{\mathrm{kin-r}}>T_{\mathrm{SD}}$), the
expansion rate follows the standard way (Eq.~\eqref{H-T relation}),
and the decoupling temperature remains unchanged. Consequently, the
abundances of elements remain unaffected. Therefore, the transition
temperature must satisfy
$T^{\mathrm{eq}}_{\mathrm{kin-r}} > T_{\mathrm{SD}} = 4\,\mathrm{MeV}$. As investigated in Ref.~\cite{Kawasaki:1999na,Kawasaki:2000en,Ichikawa:2005vw,deSalas:2015glj,Hasegawa:2019jsa,Barbieri:2025moq}, if radiation domination is realized
by entropy production through a decay of heavy particles, the effective number of neutrino species $N_{\mathrm{eff}}$ can be
changed. As clearly shown in Ref.~\cite{Hasegawa:2019jsa}, if entropy production occurs
at reheating temperatures below 4 MeV, even $N_{\mathrm{eff}}$ can be smaller than 3. Neutrino oscillations also play a role (see also
\cite{Escudero:2018mvt}), slightly increasing $N_{\mathrm{eff}}$ due to differences of
interactions among flavors. Since there is no entropy production in
the current model, this effect does not apply.  Taking this effect
into account, to ensure the success of BBN, we conservatively adopt a
lower limit where the transition temperature $T_{\mathrm{kin}}^{\mathrm{eq}}$ should be larger
than 4 MeV. We apply this condition in Sec.\ref{sec:constraint on initial
  condition} to quantify the required fine-tuning for this model.

\subsection{Constraint on Initial Condition}
\label{sec:constraint on initial condition}

In this subsection, we derive the conditions on the initial $\Omega_{\mathrm{pot}}$. We denote the e-folding number corresponding to the BBN epoch as $N_{\mathrm{BBN}}$. The potential during kination is given by
\begin{align}
    V=V_0 e^{-\lambda\phi_i/m_{\mathrm{pl}}}e^{-\lambda \sqrt{6}(N-N_i)},
\end{align}
where $N_i$ is arbitrary time during kination. Based on the discussion
in Sec.~\ref{sec:dynamics of scalar field}, the current energy scale
of the dark energy is determined at the time of the transition from
kination domination to radiation domination because the scalar field
$\phi$ is frozen from this epoch to the present time. Therefore, the
following relation holds:
\begin{align}
    V_0 e^{-\lambda \phi_i /m_{\mathrm{pl}}}e^{-\lambda\sqrt{6}(N_{\mathrm{kin-}r}^{\mathrm{eq}}-N_i)}\simeq \rho_{0},
\end{align}
where $N_{\mathrm{kin-}r}^{\mathrm{eq}}$ is the time of transition from kination to radiation domination and $\rho_0\simeq (2\,{\rm meV})^4$ is the dark energy density at the present epoch. Solving this equation for $N_{\mathrm{kin-}r}^{\mathrm{eq}}$, we obtain
\begin{align}
    N_{\mathrm{kin-}r}^{\mathrm{eq}}=N_i-\frac{1}{\lambda\sqrt{6}}\ln\left( \frac{\rho_0}{V_i} \right),
\end{align}
where $V_i\equiv V_0 e^{-\lambda \phi_i/m_{\mathrm{pl}}}$. As we
discussed in the previous section, the transition to radiation
domination must occur before $N_{\mathrm{BBN}}$. From this condition,
we obtain the upper bound on $N_{\mathrm{kin-}r}^{\mathrm{eq}}$ by
\begin{align}
    N_{\mathrm{kin-}r}^{\mathrm{eq}}=N_i-\frac{1}{\lambda\sqrt{6}}\ln\left( \frac{\rho_0}{V_i} \right)<N_{\mathrm{BBN}}.
\end{align}
Solving this inequality for $V_i$ yields the upper bound on $V_i$ to be
\begin{align}
    V_i <\rho_0 e^{\lambda\sqrt{6}(N_{\mathrm{BBN}}-N_i)}.
\end{align}
In order to estimate the degree of fine-tuning of
$\Omega_{\mathrm{pot,i}}$, by using
$V_i=\rho_i \Omega_{\mathrm{pot,i}}$, we rewrite the inequality to be
\begin{align}
    \Omega_{\mathrm{pot},i}<\frac{\rho_0}{\rho_i}e^{\lambda \sqrt{6}(N_{\mathrm{BBN}}-N_i)}.
\end{align}
Here, setting $N_i$ to the Planck scale $N_{\mathrm{pl}}\simeq-73$ and $N_{\mathrm{BBN}}\simeq-24$, we
obtain \footnote{ If we consider the observational upper bound on the
  energy scale of the primordial inflation to be the initial
  condition, $N_i$ can be set to the time of the GUT scale
  ($\sim 10^{16}$~GeV). Then, by assuming
  $N_i= N_{\mathrm{GUT}}\simeq -67$, we may obtain a possible milder
  bound than that of (\ref{eq:NiPlanck}) to be
  $\Omega_{\mathrm{pot},\mathrm{GUT}}\lesssim 10^{-112}10^{19\sqrt{6}\lambda}$.
}
\begin{align}
  \Omega_{\mathrm{pot},\mathrm{pl}} \lesssim 10^{-120} 10^{21\sqrt{6}\lambda}.
  \label{eq:NiPlanck}
\end{align}
This represents a necessary condition imposed on the initial condition
of the energy density at the Planck time. If this condition is not
satisfied, the freeze-out temperature of weak interaction at the
beginning of the BBN should have been modified and altered the
light-element abundances.  In the limit $\lambda\to 0$, this
inequality brings us back to the original problem of the fine-tuned
small cosmological constant. This result clearly indicates that the
degree of fine-tuning is alleviated when $\lambda$ is non-zero.

Note, however, that this is a necessary condition, but not necessarily
a sufficient one. This is because in the derivation of this inequality, we implicitly assumes that the kination phase as an initial condition. The derivation of the exact upper bound (i.e., the necessary and sufficient condition) is left for future work.

One might think that it is circular reasoning because we used the present value of the dark energy density in the derivation. However, this is not the case. In the standard $\Lambda\mathrm{CDM}$ model, the initial potential energy density parameter must be exactly $10^{-120}$. In contrast, our model demonstrates that to explain the current dark energy, the initial potential energy is not confined to a single fine-tuned value, but a wider region is allowed. Such an expansion of the allowed parameter region for the initial conditions is the essence of the mitigation of the fine-tuning problem in our model.

\section{Spectrum of Primordial Gravitational Waves}
\label{sec:primordial gravitational wave}
In this section, we discuss how to test this model, i.e., the
observability of the long kination epoch. It is well known that a mode
of gravitational wave background re-entering horizon during the
kination phase is significantly amplified, compared to the one during
radiation domination \cite{Tashiro:2003qp}. When the mode of
primordial gravitational wave exits the horizon, its amplitude $h_c$
is proportional to the Hubble parameter $H_{\mathrm{out}}$ at the time
of horizon crossing. Subsequently, the amplitude remains frozen until
the mode re-enters the horizon. After re-entry, it decays as
$1/a$. Hence, the amplitude can be expressed as
\begin{align}
  h_c \propto H_{\mathrm{out}}\frac{a_{\mathrm{in}}}{a}.
\end{align}
In the following analysis, we assume the constant Hubble parameter
$H_{\mathrm{out}}$ during inflation and a power-law expansion
$a \propto t^n$ after the re-entry.  Since the Hubble parameter scales
as $H \propto 1/t \propto a^{-1/n}$, the wavenumber of the
gravitational wave re-entering the horizon $k_{\mathrm{in}}$ is given
by
$k_{\mathrm{in}}=a_{\mathrm{in}}H_{\mathrm{in}}\propto
a_{\mathrm{in}}^{1-1/n}$. This implies :
\begin{align}
  a_{\mathrm{in}}\propto k^{\frac{n}{n-1}}.
\end{align}
Consequently, the energy spectrum of gravitational waves
$\Omega_{\mathrm{GW}}$ is given by
\begin{align}
  \Omega_{\mathrm{GW}}(f)=\frac{1}{\rho_c}\frac{d\rho_{\mathrm{GW}}}{d\ln f}=\frac{1}{6H_0^2}f^2h_c^2\propto f^{\frac{2n}{n-1}+2}.
\end{align}
For radiation domination ($n=1/2$), the frequency dependence of the
energy spectrum is $\Omega_{\mathrm{GW}}\propto f^0$. For kination
($n=1/3$), the spectrum rises linearly with frequency
($\Omega_{\mathrm{GW}}\propto f$). Therefore, the longer the kination
epoch, the wider the frequency band of the amplified signal.
\begin{figure}[htbp]
  \centering
  \includegraphics[width=0.8\columnwidth]{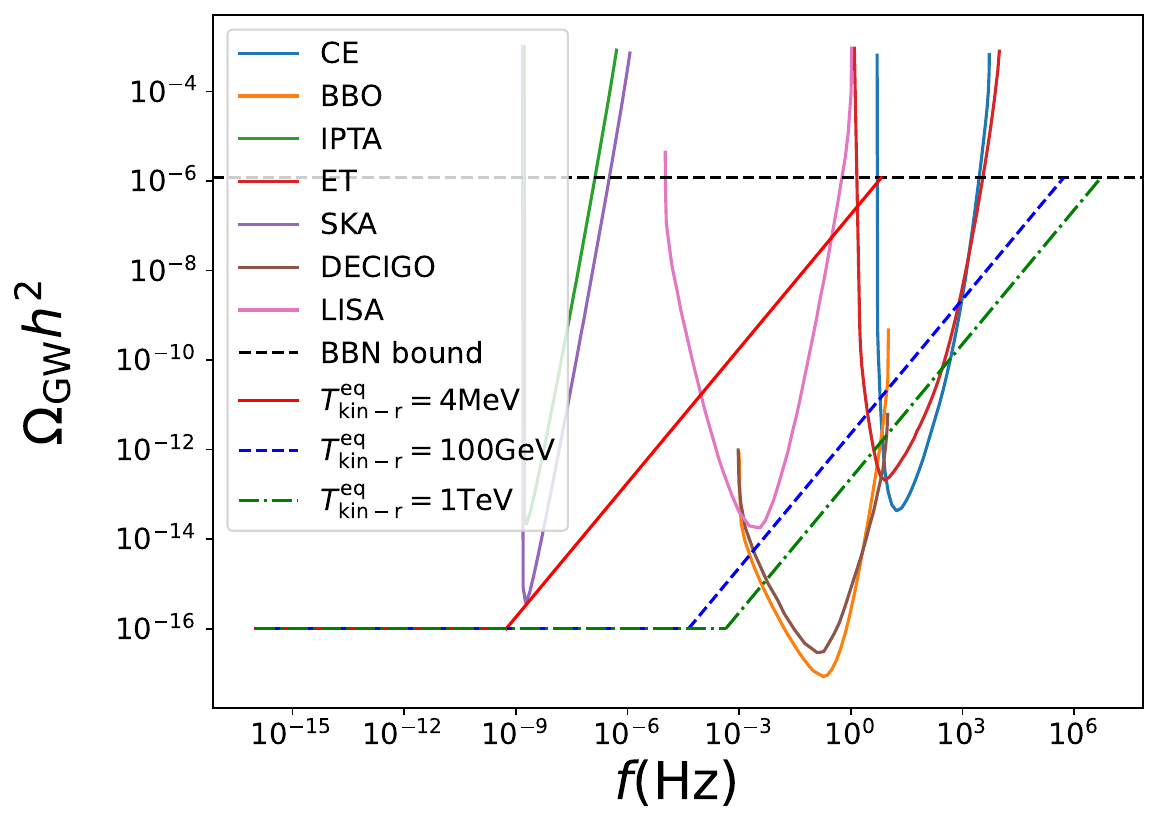}
  \caption{Energy spectra of the primordial gravitational wave
    background in case that the transition temperature from kination to
    the radiation domination is $4\,\mathrm{MeV}$, $100\,\mathrm{GeV}$,
    and $1\,\mathrm{TeV}$, respectively. The sensitivity curves of the
    gravitational wave observations are taken from Fig.~1
    of~\cite{Schmitz:2020syl}. The black dotted line represents the
    constraint $\Omega_{\mathrm{gw}}h^2<1.2\times 10^{-6}$ derived
    from combining observations of CMB, BBN, and
    BAO~\cite{Pagano:2015hma} }
  \label{fig:gravitational waves}
\end{figure}

In Fig.~\ref{fig:gravitational waves}, we plot the spectra of
gravitational wave for
$T^{\mathrm{eq}}_{\mathrm{kin}-r}=4\,\mathrm{MeV}, 100\,\mathrm{GeV}$
and $1\,\mathrm{TeV}$. The spectra are constrained by
$\Omega_{\mathrm{gw}}h^2<1.2\times 10^{-6}$ from CMB, BBN, and
BAO~\cite{Pagano:2015hma}. If the transition temperature from kination
to radiation domination is $4~\mathrm{MeV}$, the primordial
gravitational waves can be detectable by LISA, Cosmic Explorer (CE),
Einstein Telescope (ET), Big Bang Observer (BBO), DECIGO, and Square
Kilometer Array (SKA). If the transition occurs at $100\,\mathrm{GeV}$
or $1\,\mathrm{TeV}$, it would be detectable by each detector except
LISA and SKA.

\clearpage
\section{Conclusion}
\label{sec:Conclusion}
In this study, we have investigated the fine-tuning problem in the
quintessence model with the exponential potential motivated by DESI's second data release, which suggests time-varying dark energy. By requiring that the
quintessence realizes the observed current dark-energy density, and that the kination epoch ends prior to the Big Bang
Nucleosynthesis, we have derived a necessary condition for the density
parameter of the potential energy at the Planck time:
\begin{align}
  \Omega_{\mathrm{pot},\mathrm{pl}}<10^{-120}10^{21\sqrt{6}\lambda}\notag.
\end{align}
This inequality reveals a direct link to the fine-tuning problem of the cosmological constant, and it suggests that the degree
of fine-tuning is alleviated when $\lambda$ is non-zero, e.g., by
dozens of orders of magnitude for $\lambda \sim {\cal O}(1)$.

It is also worth noting that this scenario is consistent with the time-varying equation of state suggested by the recent DESI data. In addition, this model does not require fine-tuning for the timing of the variation of the equation of state parameter because this variation inevitably coincides with the beginning of dark energy domination.

Furthermore, we have discussed the detectability of the primordial
gravitational wave background amplified during the kination epoch. We
found that if the kination phase persists until the vicinity of the
BBN epoch ($T^{\mathrm{eq}}_{\mathrm{kin}-r}\simeq4\,\mathrm{MeV}$),
the gravitational wave signal could be detectable by the future
observations, SKA, LISA, BBO, DECIGO, ET or CE. Such detections would
provide new constraints on the parameters of this model.

\begin{acknowledgments}
This work was in part supported by JSPS KAKENHI Grants
Nos. JP23KF0289, JP24K07027, and MEXT KAKENHI Grants No. JP24H01825.
\end{acknowledgments}

\bibliography{reference}

\end{document}